\documentclass[prl,superscriptaddress,twocolumn,subeqnawidetrray]{revtex4}

\usepackage{graphicx}

\usepackage{amssymb}
\begin{document}
\columnsep=1cm
\frenchspacing
\baselineskip=10.5pt plus 1pt minus 0pt \par
%\noindent\begin{tabular}{lr}
\itemsep=0pt \par
\parsep=0pt \par
\parindent=7mm
\font\diezb=cmbx10
\font\diezi=cmti10
\font\diez=cmr10
\font\once=cmr10 at 10.96pt
\font\oncei=cmti10 at 10.96pt
\font\doce=cmr10 at 12pt
\font\nueve=cmr9
\font\fivei=cmmi5
\font\onceb=cmbx10 at 10.96pt
\font\catorce=cmbx10 at 14.4pt
%\font\catori=cmbxti10 at 14.4pt
%\font\diecib=cmbx10 at 17.28pt
%\font\veinteb=cmbx10 at 20.736pt
\font\doceb=cmbx10 at 12pt
\font\curna=cmbxti10 at 9pt
\font\catorb=cmbx10 at 14.4pt
\font\nueveb=cmbx9
\font\nueve=cmr9
\font\ochob=cmbx8
\font\sieten=cmbx7
\font\ocho=cmr8
\font\cinco=cmr5
\font\cincoi=cmmi5
\font\siete=cmr7
\font\seis=cmr6
\font\nuevei=cmti9
\font\ochoi=cmti8
\font\ochom=cmmi8
\font\sietei=cmmi7
\font\ochosy=cmsy8
\font\sietesy=cmsy7
\font\sevensy=cmsy7
\font\fivesy=cmsy5
\font\ochoex=cmex8
\font\tenex=cmex10
\def\ver#1#2|{\nueve #1{\siete #2}}
\def\versa#1#2|{\catorce #1{\onceb #2}}
\def\versec#1#2|{\doce #1{\nueve #2}}
\def\thepage{\nueve\arabic{page}}
\def\r{\mathbb R}
\def\n{\mathbb N}
\def\z{\mathbb Z}
\skip\footins=18pt plus 1pt minus 0pt
\def\footnoterule{\kern -3pt\hrule width .6 true in\kern 2.6pt}
\def\h{\hskip 7mm}
\def\que#1#2{\displaystyle\frac{#1}{#2}}
\def\pri{^\prime}
\let\de=\delta
\let\De=\Delta
\let\al=\alpha
\let\ga=\gamma
\def\sen{\mathop{\rm sen}\nolimits}
\let\ra=\rightarrow
\let\va=\varphi
\let\be=\beta
\let\vare=\varepsilon
\let\la=\lambda
\let\vfi=\varphi
\let\var=\varphi
\let\lam=\lambda
\let\c=\cdot
\let\ovl=\overline
\def\separa{\vskip 1pc\begin{center}
\vrule width 2cm height .5pt depth 0pt \
\vrule width 7mm height .35mm  depth .35mm \
\vrule width 2cm height .5pt depth 0pt \end{center}\vskip 2.5pc}
\let\vare=\varepsilon

\def\peque{\textfont0=\ocho\scriptfont0=\cinco\scriptscriptfont0=\cinco%
\textfont1=\ochom\scriptfont1=\fivei\scriptscriptfont1=\fivei%
\textfont2=\sevensy\scriptfont2=\fivesy%
\textfont3=\tenex\scriptfont3=\tenex\scriptscriptfont3=\tenex%
\ocho\baselineskip=9.5pt}

\let\ca=\textquotedblleft
\let\cc=\textquotedblright

\let\disp=\displaystyle
\let\ti=\times
\let\si=\sigma
\let\Om=\Omega
\let\om=\omega
\let\te=\theta
\let\c=\cdot
\let\ti=\times
\let\ov=\overrightarrow
\let\ove=\overrightarrow
\vspace*{-1cm}
\begin{center}
\parbox{16.5cm}
{\begin{center}
\baselineskip=11.5pt
\doceb
 ZERO-POINT RADIATION, INERTIA AND GRAVITATION
 \end{center}}\end{center}\vspace*{-2.5pc}
\begin{center}\parbox{16.5cm}{\begin{center}\baselineskip=24pt\diezb
R. Alvargonz\'alez y L.S. Soto\end{center}}\end{center}\vspace*{-2.5pc}
\begin{center}\parbox{16.5cm}{\begin{center}\baselineskip=24pt\diezb
Facultad de F\'{\i}sica, Universidad Complutense, 28040 Madrid, Spain\end{center}}
\end{center}\vspace*{-1pc}
\begin{center}\parbox{16.5cm}{\begin{center}\parbox{15cm}{\peque
In this paper it is shown that the forces which resist the acceleration of the
mass of the electron, $m_e$,
arising from the Compton effect, the Klein-Nishima-Kann formula for its
differential cross section and the transversal Doppler effect when the
electron moves in a straight line coincide, with $\vare<1,16\times10^{-4}$,
with the force required to propel me with the same acceleration, if the radius
of the electron is equal to its classical radius and if the forces which rise
from the interaction of the electron and zero-point radiation are equal to
those deriving from the electrostatic repulsion of the charge of the electron
against itself (Poincare's tensions). The equations worked in this paper show
that there is no difference between inertial mass and gravitational mass and
may be used to determine the value of the gravitational constant.

\vskip 6pt
PACS numbers:}\end{center}}\end{center}\vspace*{1cm}

\catcode`\@=11
\def\@oddhead{}
\def\@evenhead{}
\catcode`\@=12

\section{\nueveb INTRODUCTION}

\h Sparnaay's experiments in 1958 [4] showed the existence of zero-point
radiation which Nernst had considered as a possibility in 1916. This radiation
is inherent to space, and for that to be the case its spectral density
function must be inversely proportional to the cubes of its wavelengths. In
other words, the number of photons of wavelength $\la$ which strike a given
area during a given period must be inversely proportional to $\la^3$. In
1969, Timothy H. Boyer [4] deduced that the spectral density function of
zero-point radiation is:
$$f_\va(\la)=\que1{2\pi^2}\que1{(\la_*)^3}$$
where $\la_*$ is the number which measures the wavelength $\la$.

To this function there corres\-ponds the energy function
$$E_\va(\la)=\que1{2\pi^2}\que{hc}\la\que1{(\la_*)^3}$$

For $\la=0$, $E_\va(\la)=\infty$. There must therefore exist a threshold for
$\la$ which will henceforth be writ\-ten as $q_\la$.

In order to simplify the following arguments, it is convenient to use the
$(e,m_e,c)$
system of measurements in which
 the basic magnitudes are the quantum of
electrical charge, the mass of the electron and the speed of light. In this
system the units of length and time are, respectively, $l_e=e^2/m_ec^2$ and
$t_e=e^2/m_ec^3$. Moreover, the following results and formulae obtained in
titles [2] and [3] of the references are basic.\eject\vspace*{8cm}
\begin{itemize}
\item The flow of the zero-point photons of wavelength $nq_\la$ through an
area $(q_\la)^2$ is one photon in a lapse of time
$n^3q_\tau$ where $q_\tau=q_\la/c$.
Therefore $\left(\que{k_\la}n\right)^3$ photons of wavelenght
$nq_\la$ will flow through an area $(l_e)^2$ during any lapse of time $t_e$,
where $k_\la=l_e/q_\la=t_e/q_\tau$. See the final result of [2].
\item The minimun wavelengh in zero-point radiation is $xq_\la$. See [3], p.
7.
\item The energy flow of zero-point radiation per $(q_\la)^2$, expressed in
the $(e,m_e,c)$ system is:
$$W_x=\que{2\pi}{3\al}\que{(k_\la)^2}{x^3}\que{m_ec^2}{t_e} \mathrm{per} \
(q_\lambda)^2 , \ \ {\rm See\ [3]\ p. \ 7.}\eqno{(1)}$$ \item The
energy transferred by a photon of wavelength $nq_\la$ to an
electron is $E_{T_n}=E_n\pi[A_m]$.
\end{itemize}
$$E_n = \que{hc}{nq_\la} = \que{2\pi}\al\que{k_\la}n m_ec^2;$$
$$[A_m]=\lim_{m \ra \infty}\left[ \que{7}{12} A - \que{11}{10} A^2 + \cdots +
(-1)^{m-1} \{ m \} A^m \right]$$
$$A=\que{2\pi}\al\que{k_\la}n;$$
$$\{m\}=\que1{m+1}+\que2{m+2}-\que3{m+3}-1-\que{m(m-1)}6;$$
hence
$$E_{T_n}=\que{2\pi^2}\al \que{k_\la}n[A_m]m_ec^2.\eqno{(2)}$$
see [3], p. 8.
\begin{itemize}\itemsep=-2pt
\item The zero-point radiation transfers to the electron the energy flow
\vspace*{-4pt}
$$W_{T_x}=\que{2\pi^2}\al\que{(k_\la)^2}{x^3}[B]_m\que{m_ec^2}{t_e}\ \ {\rm
per}\ \ (q_\la)^2\eqno{(3)}\vspace*{-4pt}$$
where $B=\que{2\pi}\al\que{k_\la}x$;
\vspace*{-4pt}
$$[B_m]=\que7{48}B+\cdots+(-1)^m\que{\{m\}}{m+3}B^m\vspace*{-4pt}$$
see [3], p. 9. \item
$x^3=\que{4\pi^3(k_\la)^4(r_x)^4[B]_m}{3\al}$; where $(r_x)$ is
the radius of the electron expressed in $l_e$.

For $r_x=r_e=l_e$ we have
\vspace*{-4pt}
$$x^3=\que{4\pi^3(k_\la)^4[B]_m}{3\al}\eqno{(4)}\vspace*{-4pt}$$
see [3], p. 10.

\item $x^3=\que{2\pi^2(k_\la)^2(r_x)^2[B]_m}{3 \al G_e}$.

For $r_x=r_e=l_e$ we have
\vspace*{-4pt}
$$x^3=\que{2\pi^2(k_\la)^2[B]_m}{3 \al G_e}\eqno{(5)}\vspace*{-4pt}$$
see [3], p. 10.

From (4) and (5) we obtain
\vspace*{-4pt}
$$k_\la=\left(\que1{2\pi G_e}\right)^{1/2}\eqno{(6)}\vspace*{-4pt}$$

From here $q_\la= (2 \al \pi)^{1/2}L_P$, where $L_P$ is the
Planck's length, and
\vspace*{-4pt}
$$\begin{array}{l}
k_\la=8.143375\ti10^{20}\\[+2pt]
x=5.275601 \ti10^{27}\\[+2pt]
z=\que{k_\la}x=1.548877 \ti10^{-7}\end{array}\vspace*{-4pt}$$

The value of $k_\la$ means that the minimum wavelength of a photon is
\vspace*{-4pt}
$$q_\lambda = \frac{l_e}{k_\lambda} = 
(2 \alpha \pi)^{1/2} L_p . $$  
\end{itemize}

\catcode`\@=11
\def\@oddhead{\seis
ZERO-POINT RADIATION, INERTIA AND GRAVITATION\hfill
\siete\thepage}
\def\@evenhead{\siete\thepage\hfill{\seis
ZERO-POINT RADIATION, INERTIA AND GRAVITATION}}
\catcode`\@=12

\section{\peque \nueveb \baselineskip=11pt
ON ZERO-POINT RADIATION WITHIN A REFERENCE SYSTEM
$\hbox{\curna S}^\prime$\ MOVING
IN A STRAIGHT LINE AND WITH UNIFORM ACCELERATION, AND ON ITS INTERACTION WITH
AN ELECTRON AT REST IN $\hbox{\curna S}^\prime$}\vspace*{-4pt}
\h Let us consider a reference system $S\pri$ which coincides with the
inertial system $S_0$, until at $t=0$ it begins to move along the axis
$\overline{OX}$ at a uniform acceleration ``$a$", and which after the lapse
$t_1$ coincides with the inertial system $S_1$ which is moving at a velocity
$v_1=at_1$ relative to $S_0$. It is obvious that the zero-point radiation at
$S_1$ has the same spectral composition as at $S_0$, but this does not happen
with $S\pri$, which continues to move at a uniform acceleration ``$a$".

After a minimal lapse of time $q_\tau$, the zero-point radiation photons of
wavelength $\la$ in $S_1$, which at $t=t_1$ arrive at $P$ along a line which
makes an angle $\te$ with $\overline{O\pri X\pri}$ (see Fig. 1) are perceived
in $S\pri$ as if proceeding from a source of light $F_\va$, situated on a
line which passes through $P$ and makes an angle $\te$ with $\overline{O\pri
X\pri}$, and moving at velocity $\De v=a_qq_\tau$ relative to $O\pri$, where
$a_q$ is the acceleration ``$a$"\ expressed through a system in which the unit
of length is $q_\la$ and the unit of time is $q_\tau$. Therefore:
\vspace*{-4pt}
$$\De v=a_q\que{q_\la}{(q_\tau)^2}q_\tau=a_qc,\vspace*{-4pt}$$
which can also be written:
\vspace*{-4pt}
$$\De v=a_q\que{k_\la q_\la}{(k_\la q_\tau)^2}k_\la q_\tau=
a\que{l_e}{(t_e)^2}t_e,\vspace*{-4pt}$$
which reveals that the increase in velocity per $q_\tau$ given by an
acceleration of $a_q\que{q_\la}{(q_\tau)^2}$ is the same as the increase of
velocity per $t_e$ given by an acceleration of $a_q\que{l_e}{(t_e)^2}$; this
allows us to write:
\vspace*{-4pt}
$$\De v=at_e=a\que{l_e}{(t_e)^2}t_e=ac$$

%%%%%%%%%%%%%%%%%%%%%%%%%%% FIG ! %%%%%%%%%%%%%%%%%%%%%%%%%%%%%
\begin{figure}[h]
\centering
\resizebox{0.70\columnwidth}{!}{\includegraphics{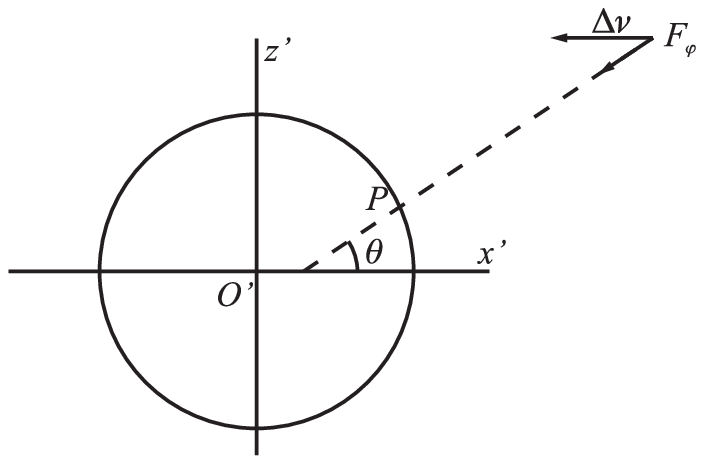}}
\caption{. }
\end{figure}

We should point out here the enormous size of the unit of acceleration in the
$(e,m_e,c)$ system, compared with the unit of the same magnitude in the
$(c.g.s.)$ system. The relation between them is given by:
\vspace*{-4pt}
$$\begin{array}{c}
\que{l_e}{(t_e)^2}=\que c{t_e}=
\que{2.997925\ti10^{10}\ {\rm cm/s}}{9.399639\ti10^{-24}\ {\rm s}}\\ \\[-8pt]
=3.189404\ti10^{33}\ \que{{\rm cm}}{({\rm s})^2}.\end{array}$$

One result of this is that the acceleration required to move in one second from
rest to the speed of light; which is the $(c.g.s)$ system is
$2.997925\ti10^{10}$ cm/(s)$^2$, is $9.399639\ti10^{-24}l_e/(t_e)^2$ when
expressed in the $(e,m_e,c)$ system. This shows that the values of ``$a$"\ are
minuscule for the accelerations which can be produced in physical reality,
excepting those which could be produced in the process of the Big-Bang. It is
therefore reasonable to conclude that ``$a$"\ is an extremely small number,
certainly less than $10^{-23}$.

Figure 1 shows photons of the zero-point radiation which, at $t=t_1$, arrive at $P$ so
as to make an angle $\te$ with $\overline{O\pri x\pri}$ and that, as
aforesaid, are perceived in $S\pri$ as if proceeding from a light-source
$F_\va$, situated on a line which passes through $P$ and moving at a
velocity $\De v=\que{al_e}{(t_e)^2}t_e=ac$ relative to $O\pri$, along a line
parallel to $\overline{O\pri X\pri}$. As a result of the Doppler lateral
relativistic effect (DLRE), their wavelength in $S\pri$ is:
$$\la\pri_0=\la\que{1-\be\cos\te}{(1-\be^2)^{1/2}};$$
where $\la$ is the wavelength of emission in $F_\va$, $\be=v/c$ and $-v$ the
velocity of the source $F_\va$ relative to $O\pri$.

By introducing in this equation $\be=\que{\De v}c=a$, we obtain
\begin{itemize}
\item For $\que\pi2>\te>-\que\pi2$; where $F_\va$ is approaching $P$

$\la\pri_\te=\la\que{1-a\cos\te}{(1-a^2)^{1/2}}$
\item For $\que{3\pi}2>\te>-\que{3\pi}2$; where $F_\va$ is moving away from
$P$
$$\la\pri_{(\pi/2+\te)}=\la\que{1+a\cos\te}{(1-a^2)^{1/2}}\eqno{(7)}$$
\end{itemize}

We have already established that the flow of zero-point radiation photons of
wavelength $nq_\la$ through a frame of area $(l_e)^2$ is one of $(k_\la/n)^3$
photons for each lapse $t_e$. Viewed from the system of reference $S\pri$,
moving at velocity $at_e$ relative to the inertial system $S_1$, and along the
axis $\overline{OX}$, this number of photons appears to pass through a frame
of area:
$$(l\pri_e)^2=\que{(l_e)^2}{(1-a^2)},$$
for every lapse $t\pri_e=\que{t_e}{(1-a^2)^{1/2}};$ this gives a photon flow
of $\left(\que{k_\la}n\right)^3(1-a^2)^{3/2}$ through a frame of area
$(l\pri_e)^2$, per $t_e$; however since the number $a$ must surely be inferior
to $10^{-23}$, this flow does not differ significantly from
$\left(\que{k_\la}n\right)^3$.

The fact that the intensity of the flow $N\pri_{n\te}$ of the photons of
wavelength $\la$ which arrive at $O\pri$ following a trajectory which makes an
angle $\te$ with $\overline{O\pri X\pri}$, through the frame of area
$(l\pri_e)^2$ during every lapse $t\pri_e$, does not differ significantly from
the intensity of the analogous flow $N_{n\te}$, enables us to establish the
value of $N\pri_{n\te}$ without difficulty. From Fig. 2 we can see that if
$N_n$ is the number of photons of wavelength $nq_\la$ which arrive from all
directions of space at a frame of area $(l_e)^2$ within a lapse $t_e$, and if
that number corresponds to the area $4\pi(N_n)^2$ of a spherical surface of
radius $N_n$, the number, $N_{n\te}$, of those which arrive at the said frame
within the same lapse of time and which make an angle $\te$ with
$\overline{OX}$, corresponds to the area produced by the arc
$\widehat{AB}=N_nd\te$ when rotated around $\overline{OX}$, whence:
$$N_{n\te}=N\pri_{n\te}=\que12\left(\que{k_\la}n\right)^3\sin\te
d\te\eqno{(8)}$$
for every $(l_e)^2$ during every $t_e$.

%%%%%%%%%%%%%%%%%%%%%%%%%%% FIG ! %%%%%%%%%%%%%%%%%%%%%%%%%%%%%
\begin{figure}[h]
\centering
\resizebox{0.70\columnwidth}{!}{\includegraphics{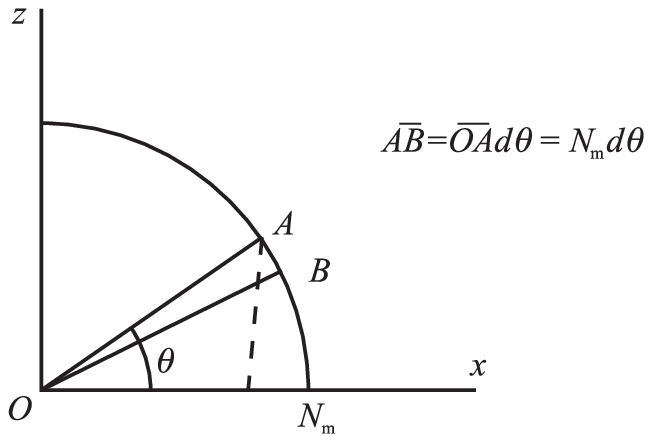}}
\caption{ . }
\end{figure}

Figure 3 shows the equatorial section of an electron at rest at $O\pri$, in
system $S\pri$. Because of the Doppler lateral relativistic effect,
the photons of wavelength $\la$ emitted by $F_\va$, moving at velocity $-a\De
v=-at_e$ relative to $O\pri$ along a straight line parallel to
$\overline{O\pri X\pri}$, and which on arrival at points $P_1$ and $P_2$ make
an angle $\te$ with $\overline{O\pri X\pri}$ are perceived at those points as
having a wavelength:
$$\la\pri_\te=\la\que{1-a\cos\te}{(1-a^2)^{1/2}}$$

%%%%%%%%%%%%%%%%%%%%%%%%%%% FIG ! %%%%%%%%%%%%%%%%%%%%%%%%%%%%%
\begin{figure}[h]
\centering
\resizebox{0.70\columnwidth}{!}{\includegraphics{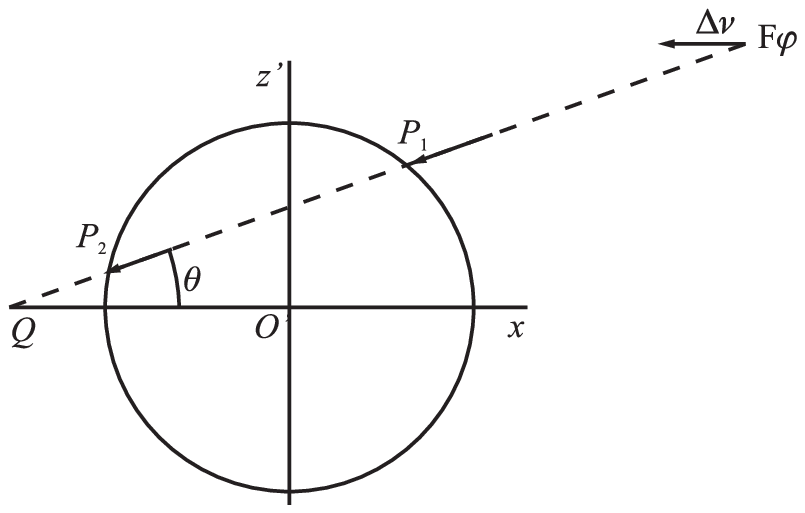}}
\caption{.}
\end{figure}

According to (8), the number of photons of wavelength $nq_\la$ which arrive at
all the points $P$ of intersection between the surface of the electron and the
cone having its apex at $Q$ and making an angle $2\te$ at that apex, is
$\que12\left(\que{k_\la}n\right)^3\sin\te d\te$.

Therefore, the energies transferred during each lapse $t_e$ to the electron by
these photons, and by those which arrive from the opposite direction,
$\pi+\te$, are respectively given by:
$$W\pri_{T_{n\te}}=\que{\pi^2}\al\left(\que{k_\la}n\right)^4
\que{(1-a^2)^{1/2}}{1-a\cos\te}\Bigl[A\pri_{n\te}\Bigr]\sin\te d\te
\que{m_ec^2}{t_e}$$
$$\!\!W\pri_{T_n(\pi+\te)}\!\!=\!\que{\pi^2}\al\!\left(\!\que{k_\la}n\!\right)^4
\!\que{(1\!-\!a^2)^{1/2}}{1\!+\!a\cos\te}\!
\Bigl[\!A\pri_{n(\pi+\te)}\!\Bigr]\sin\te d\te
\que{m_ec^2}{t_e}\!\!\!\eqno{(9)}$$
per $(l_e)^2$, where:
$$A\pri_{n\te}=- \left ( 
\que{2\pi}\al\que{k_\la}n\que{(1-a^2)^{1/2}}{1-a\cos\te} \right ) ;$$
$$A\pri_{n(\pi+\te)} = - \left ( \que{2\pi}\al\que{k_\la}n
\que{(1-a^2)^{1/2}}{1+a\cos\te} \right ) ;$$
and where the developments of $\!\Bigl[\!A\pri_{n\te}\!\Bigr]\!$ and
$\!\Bigl[\!A\pri_{n(\pi+\te)}\!\Bigl]\!$ are analogous to the development of
$[A_n]$ for $A=\que{2\pi}\al\,\que{k_\la}n$, which appears in (2).

In (9) the number to be considered in each of the formulas is the number
arriving from a half-space, so that $\pi^2/\al$ appears there instead of
$2\pi^2/\al$ as in~(2).

The total $W\pri_{T_{n\te}}+W\pri_{T_{n(\pi+\te)}}$ per $(l_e)^2$ is:
\vskip 2pc
\begin{widetext}
\begin{eqnarray*}
\que{m_ec^2}{t_e}\que{\pi^2}\al \sin\te\,d\te
\left\{ \que7{12} \left(- \frac{2\pi k_\lambda}{\alpha n}\right)^5 
\left[\que1{(1-a\cos\te)^2}-\que1{(1+a\cos\te)^{1/2}}\right]+ 
\cdots \right. \\
+ \left . \{m \} \left( - \frac{2\pi k_\lambda}{\alpha n} \right)^{m+4}
(1-a^2)^{\frac{m+1}2}
\left[\que1{(1-a\cos\te)^{m+1}}-
\que1{(1+a\cos\te)^{m+1}}\right]\right\}\left(\que\al{2\pi}\right)^4\\
=\que{m_ec^2}{t_e}\que{\pi^2}\al\sin\te d\te\left\{\que7{12}
\left(\que{2\pi k_\lambda}{\al n}\right)^5 
\que{(1-a^2)^{2/2}}{[1-(a\cos\te)^2]^2}2a\cos\te+\cdots+
(-1)^{m-1}\{m\}\left(\que{2\pi k_\lambda}{\al n}\right)^{m+4}\right.\\
\times\que{(1-a^2)^{\frac{m+1}2}}{[1-(a\cos\te)^2]^{m+1}}
\left.\left[2(m+1)a\cos\te+2\left(\begin{array}{c}
m+1\\[+3pt]
3\end{array}\right)a^3\cos^3\te+\cdots\right]\right\}
\left(\que\al{2\pi}\right)^4
\end{eqnarray*}
\end{widetext}
\noindent
Since $a\!<\!10^{-\!23}$, $(1\!-\!a^2)^{\frac{m+1}2}$ and
$\{1\!-\!(a\cos\te)^2\}^{m+1}$ do
not differ sig\-ni\-fi\-can\-tly from 1, while the terms $2\left(\begin{array}{c}
m+1\\[+2pt]
3\end{array}\right)a^3\cos^3\te$, $2\left(\begin{array}{c}
m+1\\[+2pt]
5\end{array}\right)a^5\cos^5\te$, etc., are insignificant with respect to
$2(m+1)a\cos\te$, and can be ignored so that we can write:
\arraycolsep=2pt$$\!\!\!\begin{array}{ll}
W\pri_{T_{n\te}}& +W\pri_{T_{n(\pi+\te)}}\!=\!\que{m_ec^2}{t_e}\que{\pi^2}\al
a\sin\te\cos\te d\te\\ \\
&\ti\left[\!\que7{12}\!\left(\!\que{2\pi}\al\!\right)\left(\!
\que{k_\la}n\!\right)^54\!+\!\cdots\right.\\ \\
&\!\!\!\!\!\!\left.+(-1)^{m+1}\{m\}\!\left(\!\que{2\pi}\al\!\right)^{m+4}
\!\!\!\left(\!\que{k_\la}n\!\right)^{m+4}\!\!\!
2(m\!+\!1)\!\right]\end{array}\!\!\!\eqno{(10)}$$
per\ \ $(l_e)^2$.

The energy flow which is transferred to the electron by all the zero-point
radiation photons which arrive at $P_1$ and $P_2$ with trajectories which,
respectively, make angles $\te$ and $\pi+\te$ with $\overline{O\pri X\pri}$ can be
obtained by adding up the values of $n$ between $n=x$ and $n\ra\infty$,
where $xq_\la$ is the wavelength of the zero-point radiation photon which
possesses most energy. Since $\disp\sum^\infty_{n=x}n^{-m}=
\que1{m-1}\c\que1{x^{m-1}}$, where $\varepsilon<\que1{2x}$, and $x$ is a very
large number, we can accept that the value $\que1{m-1}\c\que1{x^{m-1}}$ does
not differ sig\-ni\-fi\-can\-tly from $\disp\sum^\infty_{n=x}n^{-m}$ and by
introducing it in (10) we obtain
$$\!\!\!W\pri_{T_\te}\!=\!\que{2\pi^2}\al
k_\la\!\left(\!\que{k_\la}x\!\right)^3\!-\!\!\!\![\,B_x-\!\!\!]\
a\sin\te\cos\te d\te\que{m_ec^2}{t_e}\!\!\!\eqno{(11)}$$
per\ \ $(l_e)^2$, where
\begin{eqnarray*}
-\!\!\![\,B_x -\!\!\!] & = & -\!\!\!\!\!\left[\,\que{14}{48}(B_x)-
\que{33}{50}(B_x)^2+\cdots \right. \\
& + & \left .  (-1)^{m-1}\{m\}\que{m+1}{m+3}(B_x)^m-\!\!\!\!\right],
\end{eqnarray*}
and $B_x=\que{2\pi k_\la}{\al x}$.

Hence we obtain the force $\!F\pri_\te\!=\!\que{W\pri_{T_\te}}c\!$, whose projection
along $\ovl{O\pri X\pri}$ is
$\que{W\pri_{T_\te}\cos\te}c$; in other words:
$$\!\!F\pri_\te\cos\te\!=\!\que{2\pi^2}\al k_\la\!\left(\!\que{k_\la}x\!\right)^3
\!\!\![\,B_x-\!\!\!]\; a\cos^2\te\sin\te d\te\que{m_ec}{t_e}
\!\!\!\!\eqno{(12)}$$
per\ \ $(l_e)^2.$

Figure 4 shows that for an area $(q_\la)^2$ situated on the surface of the
electron at $P$ on $\ovl{O\pri X\pri}$, the sum of the projections
$F\pri_\te\cos\te$ is:
$$F\pri_{P_\te}=\que{2\pi^2}\al\que{(k_\la)^2}{x^3}-
\!\!\!\![\,B_x-\!\!\!]\
a\que{m_ec}{t_e}\int^{\pi/2}_0\cos^2\te\,\sin\te\,d\te,$$
per $(q_\la)^2,$
whence
$$F\pri_{P_0}=\que{2\pi^2}{3\al}\que{(k_\la)^2}{x^3}-\!\!\!\![\,B_x-\!\!\!]\
a\que{m_ec}{t_e}\ \ \ {\rm per}\ \ (q_\la)^2.\eqno{(13)}$$
where $\que{(k_\la)^2}{x^3}$ appears instead of
$k_\la\left(\que{k_\la}x\right)^3$ because (12) expresses forces per $(l_e)^2$;
while we are now talking of forces per $(q_\la)^2$, i.e. the former divided
by~$(k_\la)^2$.

%%%%%%%%%%%%%%%%%%%%%%%%%%% FIG ! %%%%%%%%%%%%%%%%%%%%%%%%%%%%%
\begin{figure}[h]
\centering
\resizebox{0.70\columnwidth}{!}{\includegraphics{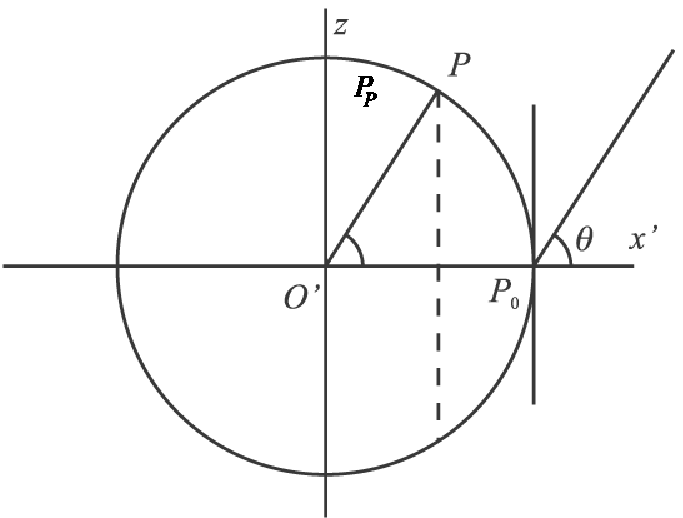}}
\caption{Fig. 1}
\end{figure}

The sum $F_P$ of the projections along $\ovl{O\pri X\pri }$ of the forces which come
from zero-point radiation and the Doppler lateral relativistic effect is the
same at every point inside the circle whose center is at $O\pri$ and whose
radius is $O\pri P_P$, perpendicular to $O\pri X\pri$.
%of the surface of the half sphere where $P_0$ is
%positioned. Notwithstanding only at $P_0$ this sum of forces strikes
%perpendiculary on the tangent plane in $P$. At any other point it strikes with
%an angle $\va$ and its intensity by $(q_\la)^2$ on that plane is $F_P\cos\va$.
%The sum of the products $F_P\cos\va(q_\la)^2$ for all the points of the said
%half sphere is equal to $F_P\cdot\pi(r_xl_e)$ where $r_xl_e$ is the length of
%the radius.
Therefore, the force which comes from the presence of zero-point
radiation and the Doppler lateral relativist effect is
$$F=\que{2\pi^3}{3\al}\que{(k_\la)^4(r_x)^2}{x^3}-\!\!\!\![\,B_x-\!\!\!]\
a\que{m_ec}{t_e}\eqno{(14)}$$
where $r_x$ is an integer. For $r_x=1$
$$F=\que{2\pi^3}{3\al}\que{(k_\la)^4}{x^3}-\!\!\!\![\,B_x-\!\!\!]\
a\que{m_ec}{t_e}\eqno{(15)}$$

By introducing in this last formula the values obtained in [3] for $k_\la$ and
$x$, which are reproduced in the Introduction we obtain
$F_0=0.999884a\que{m_ec}{t_e}$, instead of $F=a\que{m_ec}{t_e};$ the
difference is less than $1.16\times10^{-4}$.

For $a=G_e/d^2$ (15) gives:
$$F_G=\que{2\pi^3}{3\al}\que{(k_\la)^4}{x^3}-\!\!\!\![\,B_x-\!\!\!]\
\que{G_e}{d^2}\que{m_ec}{t_e}.$$
By equalising with $F_d=\que{2\pi^2}{3 \al} \frac{k_\lambda^2}{x^3} \que{[B]_m}{d^2}\que{m_ec}{t_e}$ (see [3],
formula (18)) we obtain:
$$G_e=\que{[B]_m}{\pi(k_\la)^2-\!\!\!\![\,B_x-\!\!\!]} \eqno{(16)}$$
instead of $G_e=\que1{2\pi(k_\la)^2}$ as in (6). The quotient:
$$\que{-\!\!\![\,B-\!\!\!]_x}{[B_m]}=$$
$$\!\!\!\!\!=\!
\que{(\frac{14}{48})-(\frac{33}{50})B\!+\!\cdots\!+\!
(-\!1)^{m-1}\{m\}(m\!+\!1)(m\!+\!3)^{-1}B^m}{%
(\frac7{48})-(\frac{11}{50})B+\cdots+(-1)^{m-1}\{m\}(m+3)^{-1}B^m}$$
is equal to 2 for $B\ra0$

$B=\que{2\pi}\al z$ must be $<1$ which means that
$$z<\al/2\pi=1.16\times10^{-3}$$
For $z=1.1\times10^{-3}$, very near to its upper limits, we have
$\que{-\!\!\![\,B-\!\!\!]_x}{[B_m]}=1.358$

For $k_\lambda = 8.143375 \times 10^{20}$ and $x = 5.257601 \times 10^{27}$,
we obtain $G =6.67487 \times 10^{-8}$, which means $1.0001 \times 
6.6742 \times 10^{-8}$. This last value was taken from CODATA 2002. In 
CODATA 1991 the value of $G$ was  $G= 6.67259 \times 10^{-8}$. 
The quotient CODATA2002/CODATA1991 is equal to 1.00024. The obtained
value would mean an additional correction approximately equal to half
the value of the anterior.

\section{\nueveb CONCLUSIONS}
\begin{enumerate}\itemsep=0pt
\item The results obtained through the set of equations (4), (5), (6) and (15)
show that inertia may be considered as the opposition to any change in the
motion of electrons immersed in zero-point radiation because of the lateral
relativistic Doppler effect (Therefore, there remains a difference equal to
$1.1438\times10^{-6}$\% to be explained).
\item The equations worked in this paper show that there is no
difference between inertial mass and gravitational mass.
\item With abstraction of the aforesaid difference, equation (15) constitutes
a new validation of Einstein's Theory of Special Relativity.
\item There are no gravitons. The photons of zero-point radiation are the
messengers of gravity.
\item Quantum theory plays an important role to explain gravity. The extremely
large and the extremely small are closely linked.
\item The equations (4) and (5) come from title [3] of the references and form
a system of two equations with two variables, $x$ and $k_\la$, which might be
the result of mere imagination. With equation (15) we have a system of three
equations with the same variables to explain three different phenomena,
apparentely independent. This system would have no solution if the equations
were the result of more imagination. Therefore they must be truly linked with
the physical rea\-li\-ty.
\item Formula (16) may be useful, possibly, to obtain better
values for $G_e$ and, therefore, for $G$.

\item Formula (15) expressed the resistence of zero-point radiation to a change
in the state of motion of the mass $m_e$ with acceleration ``$a$". This
resistence can be identified by calling it $F_{0a}$. Therefore we have
$$F_{0a}=\que{2\pi^3}{3\al}\que{(k_\la)^4}{x^3}![\,B_x-\!\!\!]\
a\que{m_ec}{t_e}$$
The mass $m_e$ is the mass which has been considered in the equations in
article [4], which leads to equations (4), (5) and (6) in this paper. We have
not considered any difference between gravitational mass and inertial mass.

Equation (19) in paper [4] expresses the force produced by the ``shadow
effect"\ between 2 electrons immersed in zero-point radiation, which is later
identified with the gravitational attraction between them, which leads to
equations (5) and (6) of this paper. If we call this force $F_{0G}$, we can
write:
$$F_{0G}=\que{2\pi^2}\al\que{(k_\la)^2}{x^3}\que{[B]_m}{d^2}\que{m_ec}{t_e};$$
whence, by using the remaining equations, we arrive at:
$$\que{F_{0a}}{F_{0G}}=\que{-\!\!\!\![\,B_x-\!\!\!]}{2[B]_m}=$$
$$\hspace*{-9mm}\!\!\!\!\!=\!\que{(\frac{14}{48})B\!-\!(\frac{33}{50})B^2\!\!+\!\cdots\!+\!(-\!1)^{m-1}
\{m\}(m\!+\!3)^{-1}\!(m\!+\!1)B^m}{2\{(\frac{7}{48})B\!-\!(\frac{11}{50})B^2\!\!+\!
\cdots\!+\!(-\!1)^{m+1}\{m\}(m\!+\!3)^{-1}B^m\}}$$
We get the same result by considering the electrostatic repulsion between two
electrons, instead of gravitation
$$F_e=\que{e^2}{(dl_e)^2}=\que1{d^2}m_e\cdot\que{l_e}{(t_e)^2},$$
where $d.l_e$ is the distance between their centres.
$$\que{F_{0a}}{F_e}=\que{2\pi^3}{3\al}\que{(k_\la)^4}{x^3}
-\!\!\!\![\,B_x-\!\!\!]\ \que1{d^2}\que{m_el_e}{(t_e)^2}\hbox{\catorce /}$$
$$\que1{d^2}\que{m_el_e}{(t_e)^2}=\que{2\pi^3}{3\al}\que{(k_\la)^4}{x^3}
-\!\!\!\![\,B_x-\!\!\!]$$
By introducing the equation (4), $x^3=4\pi^3(k_\la)^4[B]_m/3\al$, we obtain
$$\que{F_{0a}}{F_e}=\que{-\!\!\!\![\,B_x-\!\!\!]}{2[B]_m}.$$
There is no question of inertial, gravitational or electrostatic mass, but of
forces against the resistance of zero-point radiation to the change of motion
of masses. If we measure any of them considering the acceleration which
produces in a known mass and the equation $F_a = m  a$, we obtain the value
$1.0001006F_R$, being $F_R$ the force really necessary to produce the
acceleration ``$a$"\ in the mass $m$.
\end{enumerate}

\end{document}